**Mechanically reinforced MgB$_2$ wires and tapes with high transport currents**

R.Nast, S.I.Schlachter, S.Zimmer, H.Reiner, W.Goldacker

Forschungszentrum Karlsruhe, Institut für Technische Physik, P.O.Box 3640, D-76021 Karlsruhe, Germany

**Abstract**

Monofilamentary MgB$_2$-wires with a 2- or 3-component sheath containing mechanical reinforcing stainless steel (SS) were prepared and characterized. In direct contact to the superconductor Nb, Ta or Fe was used. For a selection of samples with a Fe and Fe/SS sheath, we investigated the transport critical current behaviour in magnetic fields changing systematically the geometrical shape from a round wire to a flat tape. A strong increase of the current densities in flat tapes was observed and possible reasons for this are discussed. Reinforcing the sheath in the outer layer with different amounts of stainless steel leads to a systematic field dependent decrease of the transport critical current density with increasing steel amount. This is an indication for a pre-stress induced degradation of the critical currents in MgB$_2$ wires and first $I_c$-stress-strain experiments seem to confirm this observation and interpretation.

Corresponding author:

Wilfried Goldacker

Forschungszentrum Karlsruhe, Institut für Technische Physik

P.O.Box 3640, D-76021 Karlsruhe, Germany

Tel: +49-7247-824179

Fax: +49-7247-825398

e-mail: wilfried.goldacker@itp.fzk.de

## 1. Introduction

Since the discovery of superconductivity in $MgB_2$ [1], several authors already reported on first wires and tapes with very high transport critical current densities measured up to about $2 \cdot 10^5$ Acm$^{-2}$ [2-5] and mechanically reinforced sheath composite [5]. Tapes showed significant higher $J_c$ values than wires, probably a consequence of a higher deformation pressure with improved phase densification during rolling and possibly the presence of some phase texture. Above a certain level of about $2 \cdot 10^5$ Acm$^{-2}$, transport currents could not be measured due to thermal instability of the conductors, but extrapolations in comparison to magnetic measurements outline a potential of $J_c = 10^6$ Acm$^{-2}$ (4.2K, self field) and more. It was found that favourably a heat treatment of the final conductor with phase decomposition and reformation should be applied to improve the filament densification. This, however, restricts the choice of the sheath material in contact with $MgB_2$ due to chemical reasons. Ta and Nb tend out to be suitable showing only minor reaction layers, while Fe, being ferromagnetic, does not react with $MgB_2$. It was shown that for Nb and Ta a mechanical steel reinforcement is necessary to induce stabilising filament pre-compression which is in general recommended for mechanically stable wires [5]. The goal of the presented work was a systematic investigation of the behaviour of the transport currents with conductor geometry and the influence of varied amounts of steel reinforcement on filament pre-compression and superconducting properties.

## 2. Experimental

To reduce contact resistivity, which is a source of thermal heating at high currents, a sample series with Fe tubes (10 mm diam., 1.5 mm w.th.) filled with commercial $MgB_2$ (ALFA) was prepared. In the first series a round wire deformed by drawing, a square wire deformed by progressive turk head deformation (4-side-rolling) and four tapes (sample B1-B4) with cross section aspect ratios up to 9 were prepared (fig.1). The round wire was reinforced by outside stainless steel tubes applying at approx. 1.5 mm diameter, followed by further drawing steps down to 1 mm diameter. The final heat treatment was performed as given in ref. [5]. The three different applied steel tubes led to steel contents of 37, 54 and 66 % (fig.2). Transport critical currents were measured with standard 4-point method at 4.2 K and B = 0-10 T using a 1 µV/cm criterion.

## 3. Results

In fig.3 the transport critical current densities for the wires and tapes shown in fig.1 are given. In contrast to magnetisation measurements, the current density vs. field graph changes to a much smaller slope below approx. 5.5 T and consequently also to much lower critical current densities. $U(I)$ curves show a sudden take off in this regime with danger of tape overheating and melting, which indicates an non sufficient thermal stabilisation at such high currents. In some cases the $I_c$ transition was depending on the current sweep ramp, with slightly higher currents for fast sweep ramps, which support this interpretation. Since this observation is also found for completely absent contact resistance, creation of internal hot spots in the filaments may be an important source too. The investigation of better stabilized multifilamentary conductors may bring clarification to this observation. At fields above 5 T a continuous increase of the current densities with tape aspect ratio and flattening of the $d(\log(J_c))/dB$ slope was found. Fig.4 illustrates that nearly the threefold current density is obtained (at $B = 6$ T) changing from a wire to a tape geometry. A further significant increase seems possible from this graph. The measured $J_c$ of 27000 Acm$^{-2}$ at 4.5 T of tape B4 is comparable to the highest reported values as given in ref.[4]. The quadratic wire was prepared in a turk head to apply, like in the tapes, high rolling pressure from four sides to a wire geometry. The improved value of the quadratic wire, however, is still much lower compared to the tapes and too low to attribute the $J_c$ increase with aspect ratio in tapes completely to a better filament densification upon rolling. X ray pole figure experiments show the presence of some broad texture in the tapes which could be a further possible reason for the larger $J_c$ values [6].

The $J_c$ values of the steel reinforced wires of fig.2 are shown in fig.5. For increasing steel content $J_c$ decreases, the decrease becoming larger towards higher background fields. Going from 54 to 66 % steel content the effect nearly saturates. This behaviour resembles a typical pre-stress induced effect as very well known from investigations on $Nb_3Sn$ wires. First $J_c$ vs. axial stress/strain measurements show an increase of $J_c$ with strain which supports this interpretation, more systematic experiments are in work.

## 4. Conclusions

We found a strong increase of $J_c$ with aspect ratio for $MgB_2$/Fe tapes. Improved filament densification alone obviously could not explain this effect. A weakly textured filament could not be proved to be responsible so far, but might be an explanation and will be decided by systematic ongoing investigations. Stainless steel reinforcements very effectively stabilized the conductor but caused typical pre-stress induced $J_c$ degradations. The current limitations at low or zero fields seem to be caused by thermal heating effects, obviously in the filament itself. Reasons may be porosity, inclusions (secondary phases) and inhomogeneities. Changing to multifilamentary wires may reduce this problem, but further improvements of the $MgB_2$ powder quality and homogeneity are of high importance in parallel.

# Acknowledgement

G. Linker contributed to these investigation with first very valuable X ray pole figure spectra.

**Figure captions**

Fig.1   Cross sections of annealed $MgB_2$/Fe wires and tapes. The scale is 1.1 x 1.1 mm for the quadratic wire (top right). Tapes are named B1-B4 from top to bottom.

Fig.2   Cross sections of $MgB_2$/Fe and three $MgB_2$/Fe/SS wires with 37, 54, 66% steel content, the wire diameter being 1 mm.

Fig.3   Transport critical current densities for $MgB_2$/Fe wire and tapes (see Fig.1) at T = 4.2 K

Fig.4   Transport critical current density vs. cross section aspect ratio for $MgB_2$/Fe wires and tapes (see Fig. 3) at $B$ = 6 T, $T$ = 4.2 K

Fig.5   Transport critical current densities for a $MgB_2$/Fe wire and three $MgB_2$/Fe/SS wires (see Fig.2).

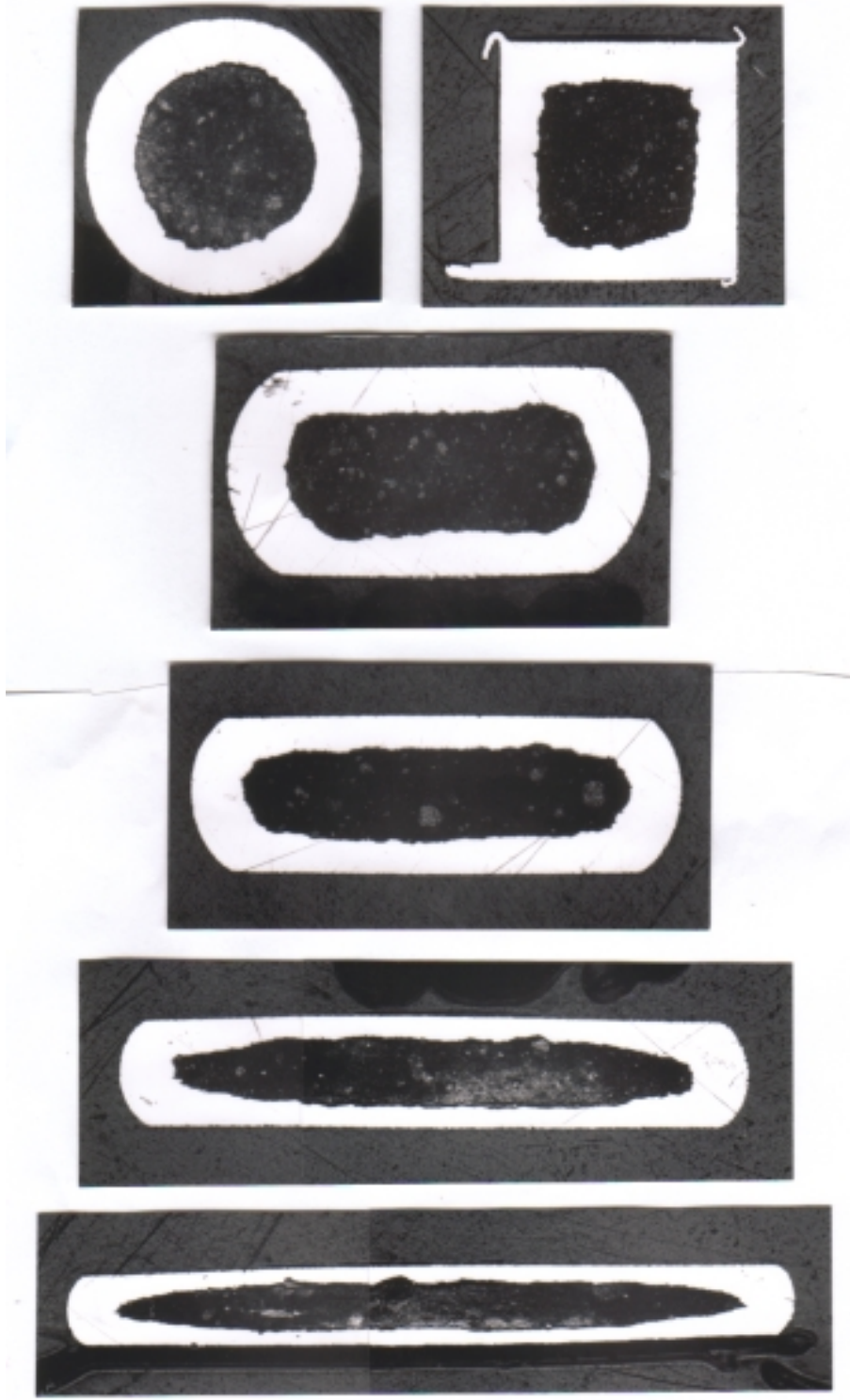

Fig.1

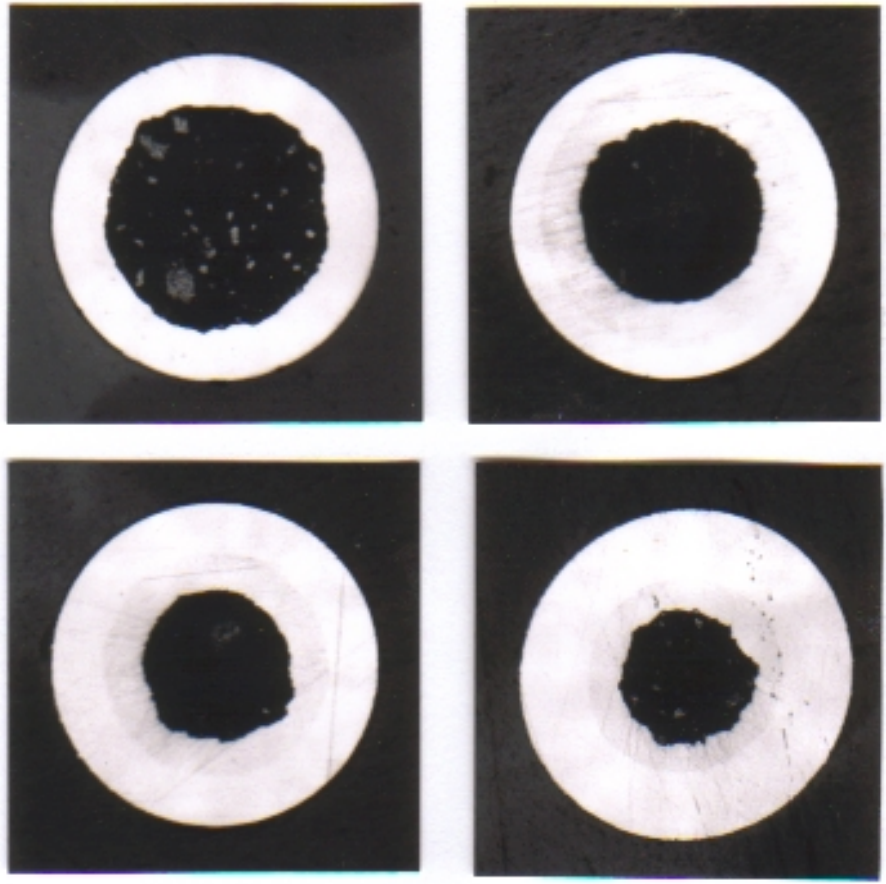

Fig. 2.

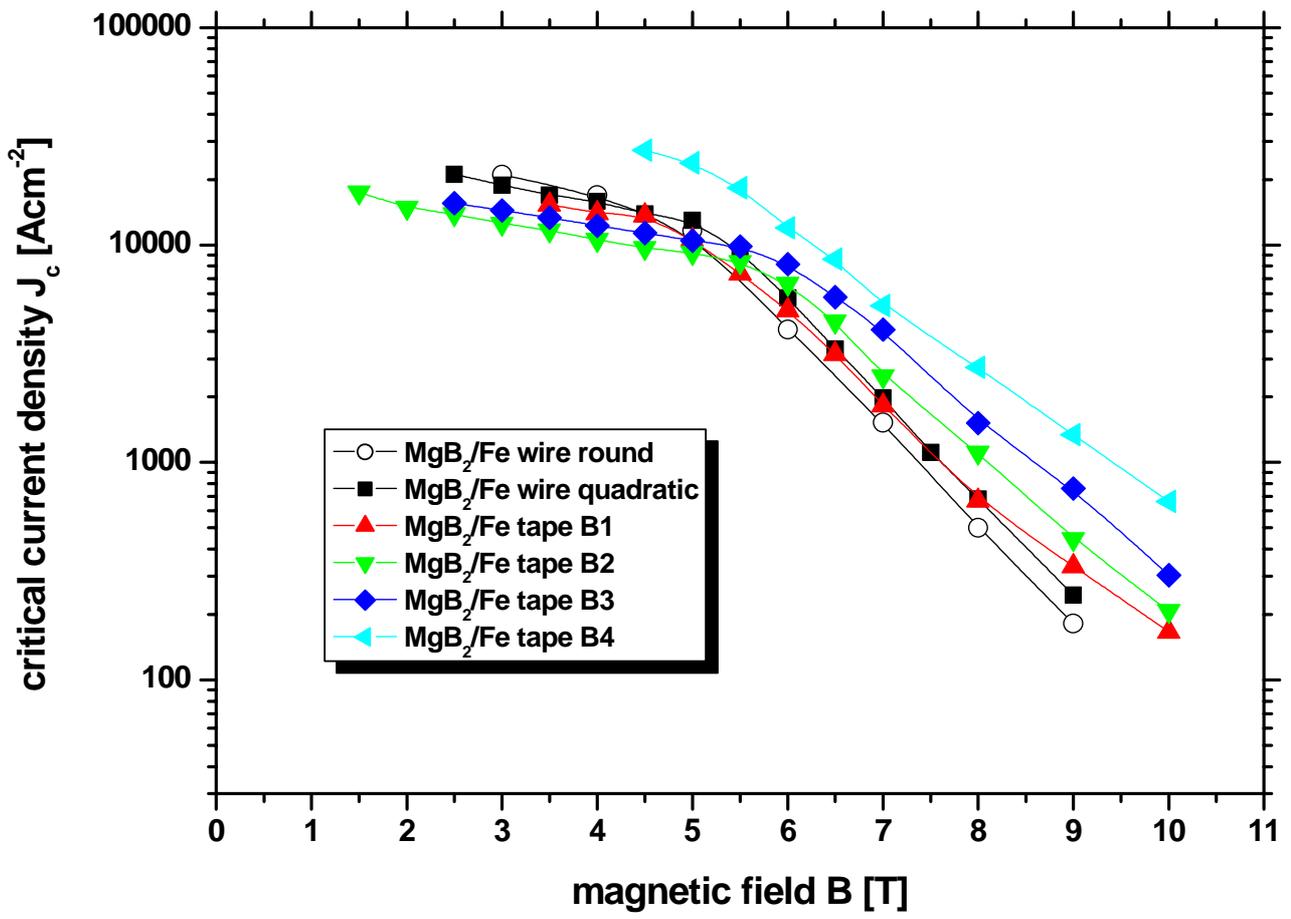

Fig. 3.

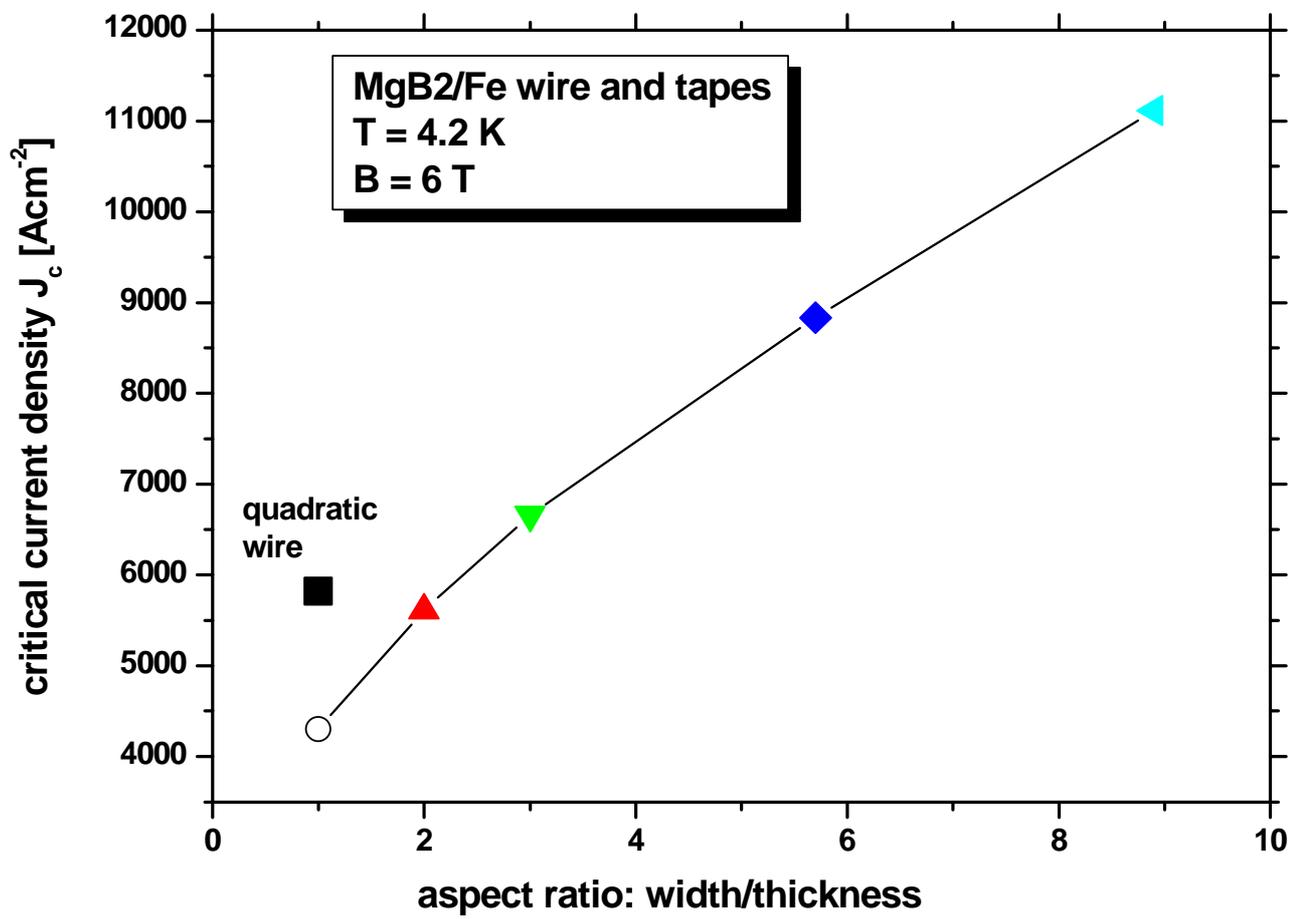

Fig. 4.

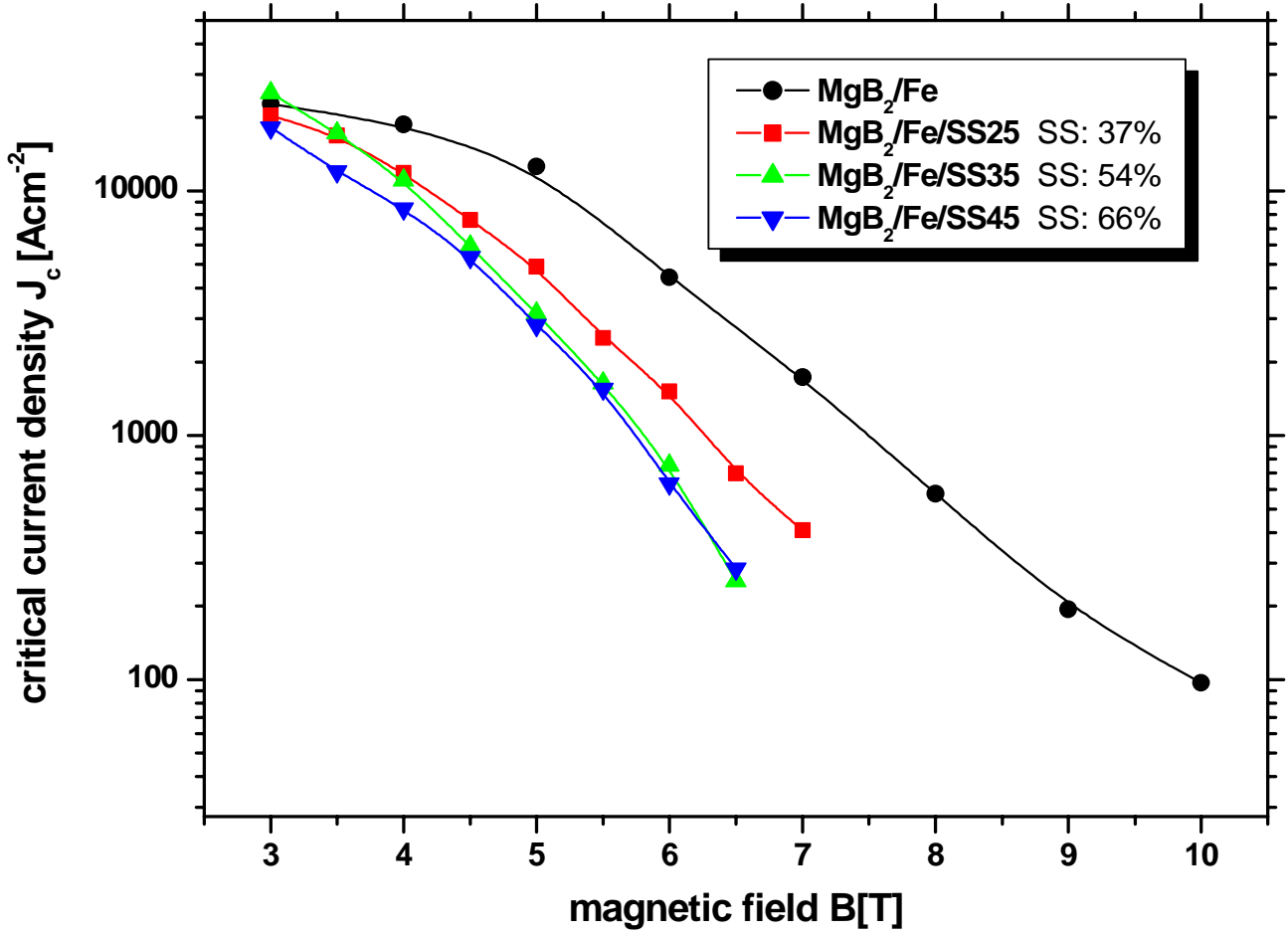

Fig. 5.